\documentclass[12pt]{article}

\pdfoutput=1

\usepackage[pdftex]{graphicx}
\usepackage{amssymb}

\newcommand{\rf}[1]{(\ref{#1})}
\newcommand{\beq}{\begin{equation}}
\newcommand{\eeq}{\end{equation}}
\newcommand{\bea}{\begin{eqnarray}}
\newcommand{\eea}{\end{eqnarray}}

\newcommand{\e}{\mbox{e}}

\newcommand{\lam}{\lambda}

\renewcommand{\a}{\alpha}

\newcommand{\del}{\delta}

\newcommand{\dg}{\dagger}

\newcommand{\ra}{\rangle}
\newcommand{\la}{\langle}
\newcommand{\prt}{\partial}

\newcommand{\cH}{{\cal H}}

\newcommand{\hH}{{\hat{H}}}

\newcommand{\cuum}{|{\rm vac}\rangle}

\newcommand{\GG}{G}
\newcommand{\pp}{p}
\newcommand{\qq}{q}

\begin{document}

\begin{center}
\vspace{24pt}
{ \Large \bf A model for emergence of space and time}

\vspace{24pt}

{\sl J.\ Ambj\o rn}$\,^{a,b}$,
and {\sl Y.\ Watabiki}$\,^{c}$

\vspace{10pt}

{\small

$^a$~The Niels Bohr Institute, Copenhagen University\\
Blegdamsvej 17, DK-2100 Copenhagen \O , Denmark.\\
email: ambjorn@nbi.dk
\vspace{10pt}

$^b$~Institute for Mathematics, Astrophysics and Particle Physics
(IMAPP)\\ Radbaud University Nijmegen, Heyendaalseweg 135, 6525 AJ, \\
Nijmegen, The Netherlands

\vspace{10pt}

$^c$~Tokyo Institute of Technology,\\ 
Dept. of Physics, High Energy Theory Group,\\ 
2-12-1 Oh-okayama, Meguro-ku, Tokyo 152-8551, Japan\\
{email: watabiki@th.phys.titech.ac.jp}

}

\end{center}

\vspace{24pt}

\begin{center}
{\bf Abstract}
\end{center}

We study  string field theory (third quantization) of the two-dimensional
model of quantum geometry called generalized CDT 
(``causal dynamical triangulations''). 
Like in standard non-critical string theory the
so-called  string field Hamiltonian of generalized
CDT can be associated with W-algebra generators
through the string mode expansion.
This allows us to define an ``absolute'' vacuum.
``Physical'' vacua appear as coherent states
created by vertex operators acting on the absolute vacuum. 
Each coherent state corresponds to specific values of 
the coupling constants of generalized CDT. 
The cosmological ``time'' only exists relatively to a given 
``physical'' vacuum and comes into existence before space, which is created
because the ``physical'' vacuum is unstable. Thus each CDT ``universe''
is created as a ``Big Bang'' from the absolute vacuum, its time evolution
is governed by the CDT string field Hamiltonian with given coupling constants,
and one can imagine interactions between CDT universes with different coupling
constants (``fourth quantization'')


\newpage

\section{Introduction}\label{intro}

Two-dimensional models  can be useful when 
it comes to addressing a number of conceptual issues related
to the quantization of geometry, simply because the corresponding 
quantum field theory is well defined and  explicit 
calculations can be performed. Here we will consider the 
model of quantum geometry denoted ``causal dynamical triangulations''
(CDT) \cite{al}. The name refers to the regularization of the continuum 
theory, which is regularized by triangulating spacetime in a specific
way using the path integral formalism. The continuum limit is obtained 
when the cut off, the link length $a$ used in the triangulations, is removed.
This limit is well defined and corresponds to  quantized two-dimensional 
Ho\v{r}ava-Lifshitz gravity \footnote{CDT can be 
formulated also in higher dimensions, and 
also in that case there  is seemingly a continuum limit of the regulated theory
(see \cite{higherCDT} for the original articles, \cite{physrep} for a 
recent review)} (for Ho\v{r}ava-Lifshitz gravity see \cite{horava}, 
for the CDT connection see \cite{agsw}).

In two-dimensional CDT it is assumed that space, which is  one-dimensional, 
has the topology of a circle.  CDT then describes the quantum 
``propagation'' of space as a function of time. Here we will 
consider so-called generalized CDT (GCDT) where one allows space to 
split and join into disconnected circles as a function of time 
\cite{GCDT}. A complete ``string field theory'' which allows us to 
calculate any  such amplitude has been developed \cite{GSFT}. 
It is inspired by the 
string field theory for non-critical string theory \cite{NCSFT,watabiki,aw}. 
Both theories are perturbative theories in the topology of the space-time 
connecting the ``incoming'' (``initial'') spatial boundaries and the 
``outgoing'' (``final'') spatial boundaries \footnote{It is even 
possible to perform certain sums over all topologies, both 
in the case of non-critical string field theories \cite{genus-sft} and
in the case of GCDT \cite{sumgenus}.}. 

In the case of non-critical string field theory $W$-algebras play an
important role and are intimately related to integrable KP hierarchies
associated with non-critical string theories \cite{kdv,aw}. In the case 
of GCDT this relation is not yet fully developed, but most likely it
exists. Multicritical GCDT and Ising models coupled to GCDT can be 
formulated \cite{moreGCDT} and the associated $W$-algebras can be 
identified \cite{toappear}. However, here we will concentrate on the 
very simplest GCDT model, its associated $W$-algebra and a possible
physical interpretation. In  Sec.\  \ref{sec2} we show how the 
$W$-algebra appears in GCDT and we discuss how the 
simplest $W$-Hamiltonian, being a 
Hamiltonian with no coupling constants and no space-time 
interpretation, contains the string field theory of GCDT 
and the seeds for a Big Bang. 
Sec.\ \ref{discuss} contains conclusion and discussion.

\section{The {\it W}-  and GCDT Hamiltonians}\label{sec2}

The formal definition of a $W^{(3)}$ algebra in terms of 
operators $\a_n$ satisfying   
\beq\label{jx10} 
[\a_m,\a_n]= m\; \del_{0,n+m}.
\eeq
is the following
\beq\label{jx11}
\a(z) = \sum_{n\in Z} \frac{\a_n}{z^{n+1}},~~~~
W^{(3)}(z)= \;\frac{1}{3}:\a(z)^3: \;= \sum_{n\in Z} \frac{W^{(3)}_n}{z^{n+3}}.
\eeq
The normal ordering $:\!(\cdot)\!:$ 
refers to the $\a_n$ operators ($\a_n$ to the 
left of $\a_m$ for $n>m$)~\footnote{We remark that this ordering is 
opposite to the standard ordering one would use in conformal field
theory. One can obtain the conventional ordering by the so-called
$\star$-operation \cite{aw}, where one uses  generating functionals
like  \rf{jy4} to express the action of the $\alpha_n$ operators
by the  action of differential operators acting on the sources of the 
generating functional. That is also the more precise way   
the $W^{(3)}$ algebra becomes related to integrable KP-hierarchies 
in non-critical string theory: certain Dyson-Schwinger equations 
satisfied by the generating functionals combined with the $W^{(3)}$-algebra
properties of the $\alpha_n$'s represented as differential 
operators acting on the sources ensure integrability 
(see i.e.\ \cite{aw} for details for 
non-critical string field theory and \cite{toappear} for GCDT string field
theory). Here we omit for transparency these technical details.}  
and we have
\beq\label{jx12}
W^{(3)}_n = \frac{1}{3} \sum_{a+b+c=n} :\a_a\a_b\a_c:.
\eeq  
In the $W^{(3)}$-algebra related to non-critical string field theory 
$\a_0$ is identical zero (see \cite{aw} for details), but in 
the GCDT case $\a_0$ plays a special role and we thus write 
\begin{eqnarray}\label{jy1}
\alpha_n =
\left\{
\begin{array}{cl}
\displaystyle
a_n^\dagger 
& \hbox{[\,$n \!>\! 0$\,]}
\rule[-12pt]{0pt}{30pt}\\
\displaystyle
\pp 
& \hbox{[\,$n \!=\! 0$\,]}
\rule[-12pt]{0pt}{30pt}\\
\displaystyle
- n  a_{-n} 
& \hbox{[\,$n \!<\! 0$\,]}
\end{array}
\right.
\end{eqnarray}
where the operators satisfy
\beq\label{js1}
[\, a_m \,,\, a_n^\dagger \,]
=
\delta_{m,n}
\qquad [\, a_m,\,a_n\,]=[\, a_m^\dg,\,a_n^\dg\,]=0
\eeq
\beq
[\, \qq \,,\, \pp \,]=i,\qquad [\, \qq \,,\, \qq \,]=
[\, \pp \,,\, \pp \,]=0 \label{js2}
\eeq
\beq
[\, \pp \,,\, a_n^\dagger \,]=[\, \pp \,,\, a_n \,]
=[\, \qq \,,\, a_n^\dagger \,]=[\, \qq \,,\, a_n \,]=0. \label{js3}
\eeq
In \rf{js2} and \rf{js3} we have introduced an operator 
$q$ conjugate to $p=\alpha_0$.
We then define the ``absolute vacuum'' $|0\ra$ by the following condition:
\begin{equation}\label{jy2}
a_n |0\rangle = \pp |0\rangle = 0
\qquad\quad [\,n =  1,2, \ldots\,],
\end{equation}
and the so-called $W$-Hamiltonian $\hH_W$: 
\begin{eqnarray}\label{jy3}
{\hH}_{\rm W} &:=& - W^{(3)}_{-2} \\
&=&-\sum_{\stackrel{\scriptstyle n,\,m,\,l}
               {[n+m+2 = l\,]}}
  a_n^\dagger a_m^\dagger l a_l
\,-
\sum_{\stackrel{\scriptstyle n,\,m,\,l}
               {[n+2 = m+l\,]}}
  a_n^\dagger m a_m l a_l\nonumber\\
&&-2\sum_{\stackrel{\scriptstyle n,\,l}
               {[n+2 = l\,]}}
  p\!\> a_n^\dagger l a_l
-p\!\> a_1 a_1
-2 p^2 a_2. \nonumber
\end{eqnarray}
Note that $\hH_W$ does not contain any coupling constants.

Related to  $\hH_W$ and the  absolute vacuum we now 
define a generating functional  with sources $x,y$
\begin{equation}\label{jy4}
Z[x,y;T] :=
\langle 0 |
  \exp\!\bigg( \sum_{n=1}^\infty y_n a_n\bigg)
  e^{- T {\hH}_{\rm W}}
  \exp\!\bigg( \sum_{n=1}^\infty x_n a_n^\dagger\bigg)
| 0 \rangle
\end{equation}
The states in the Hilbert space $\cH$ associated with $\hH_W$ are obtained
by acting repeatedly on the absolute vacuum $|0\ra$ 
with the operators $a_n^\dg$ and $q$. Such  an ``initial'' state is then 
``propagated'' a ``time'' $T$ and projected onto a similar ``final'' state.
These amplitudes can be obtained from the generating functional 
$Z[x,y;T]$ by differentiation with respect to  $x$ and $y$.     
However, we should stress that at this point there is no compelling
reason to denote $T$ a (Euclidean) time and the form of $\hH_W$ does
not suggest any obvious geometry-interpretation. One could equally
well view $T$ as an ``inverse temperature'' and use $Z[x,y;T]$ to
calculate the corresponding partition function. Here we will 
view the states and dynamics associated with $\hH_W$ as ``pre-geometry'',
and only by a projection onto a subspace of $\cH$ the parameter 
$T$ will  get an 
interpretation as (Euclidean) time and the states will obtain 
an interpretation as spatial geometries, and the amplitudes will
then be probability amplitudes for the propagation of spatial 
geometries in (Euclidean) time.  This reinterpretation of $\hH_W$
will be made by relating it to  the standard string field 
Hamiltonian $\hH$ of GCDT defined relatively to a ``physical''
vacuum $\cuum$. 

Recall the following representation of the GCDT $\hH$ 
(the one originally used in \cite{GSFT}): 
\bea\label{s8}
\hH &=& \hH_0 - g \int dL_1 \int dL_2 \;\Psi^\dg(L_1)\Psi^\dg(L_2)\;
(L_1+L_2)\Psi(L_1+L_2)
\\ && -  gG\int dL_1 \int dL_2 \;\Psi^\dg(L_1+L_2)\;L_2\Psi(L_2)\;L_1\Psi(L_1)
-\int dL \; \rho(L) \Psi(L), \nonumber
\eea
where 
\beq\label{s6}
\hH_0 = \int_0^\infty dL \; \Psi^\dg (L) 
H_0  \Psi(L),~~~~~H_0= -  \frac{\prt^2}{\prt L^2}\, L
+\mu L,~~~~\rho(L)= \del(L),
\eeq
and where the operators $\Psi(L)$ and $\Psi^\dg(L)$ satisfy 
\beq\label{jz10}
[\Psi(L),\Psi^\dg(L')]=\del(L-L'),~~~~\Psi(L) \cuum =0.
\eeq 
In \rf{s8} $\Psi^\dg(L)$ creates a spatial universe of length $L$ from the 
physical vacuum $\cuum$. The vectors $|L\ra= \Psi^\dg(L) \cuum$, $L$ positive,
span the Hilbert space where $H_0$ is defined (see \cite{GSFT} for details).
$\hH$ represents a third quantization in the sense that space can be 
created from the vacuum $\cuum$ by acting with $\Psi^\dg(L)$ and 
annihilated by acting with $\Psi(L)$.
Thus $\hH_0$ propagates spatial slices in time, can change their lengths
but cannot merge or split the spatial splices. $\mu$ denotes the 
cosmological constant and acts to limit the growth of the spatial 
universe. The second term on the 
rhs of \rf{s8} splits a spatial slice of length $L_1+L_2$ in two slices
of lengths $L_1$ and $L_2$, governed by a coupling 
constant $g$ of mass dimension 3. The third term on the rhs of \rf{s8} merges
two spatial slices of length $L_1$ and $L_2$ into one slice of length
$L_1+L_2$, governed by a coupling constant $g\cdot G$, where $G$ is 
dimensionless and is introduced to allow for a potential asymmetry between 
splitting and joining. 
Finally the fourth term on the rhs of \rf{s8} is a tadpole 
term which allows a spatial slice to disappear into the vacuum, but
only if its length is zero. Thus the interaction terms in $\hH$ 
preserve the total length of the spatial slices and any expansion or 
contraction of the universe is caused by $\hH_0$ and the coupling constant
for topology change of spacetime is $g^2G$. $\hH$ is not Hermitian
because of the tadpole term (and also if $G \neq 1$), but that is always
the case for non-critical string field theory and is enforced upon us
by the requirement of stability of the vacuum.

We now make a so-called {\it mode expansion} of $\hH$. 
The modes $\phi_n$, $\phi_n^\dg$ are defined as follows
\beq\label{j3}
\Psi (-\zeta) = \sum_{n=1}^\infty  \phi_n\,\zeta^n,~~~~
\Psi^\dg (\zeta) = \frac{1}{\zeta}+
\sum_{n=1}^\infty \frac{\phi^\dg_n}{\zeta^{n+1}}.
\eeq
where 
\beq\label{j1}
\Psi^\dg( \zeta) = \int_0^\infty dL \; e^{-\zeta L} \Psi^\dg(L),
\eeq
and similar for $\Psi$. By construction we have  
\beq\label{jz6}
\phi_n \cuum =0, ~~~~[\phi_n,\phi^\dg_m]= \del_{n,m}.
\eeq
and after some algebra (see \cite{toappear} for more details
and mode expansions also for GCDT coupled to matter) we obtain
\begin{eqnarray}\label{jx3}
\hH
\ &=& \ 
\mu \phi_1
- 2 g \phi_2
- g \GG \phi_1 \phi_1 
-\> \sum_{l=1}^\infty \phi_{l+1}^\dagger l \phi_l
+ \mu \sum_{l=2}^\infty \phi_{l-1}^\dagger l \phi_l
- 2 g \sum_{l=3}^\infty \phi_{l-2}^\dagger l \phi_l 
\nonumber \\
&&
-\> g
  \sum_{l=4}^\infty \sum_{n=1}^{l-3}
  \phi_n^\dagger \phi_{l-n-2}^\dagger
  l \phi_l
-\> g \GG
  \sum_{l=1}^\infty \sum_{m=\max(3-l,1)}^\infty
  \phi_{m+l-2}^\dagger
  m \phi_m l \phi_l.
\end{eqnarray}

Let us now relate the physical vacuum $\cuum$ to the  absolute 
vacuum $|0\ra$ and $\hH_W$ to $\hH$. 
We define the physical vacuum as the following coherent state 
relative to the absolute vacuum:
\beq\label{jz1a}
|\nu\ra =\e^{i\nu q} |0\ra,~~~ \cuum_{\nu} \,=\, 
V(\lam_1,\lam_3)\;|\nu\ra,
\eeq
\beq\label{jz1}
V(\lam_1,\lam_3):=\,
\exp\!\bigg(\! - \frac{|\lambda_1|^2}{2} - \frac{|\lambda_3|^2}{2}
  + \lambda_1 a_1^\dagger + \lambda_3 a_3^\dagger \bigg)
\eeq
and we have 
\begin{eqnarray}\label{jz2}
a_1 \cuum_{\nu} = \lambda_1 \cuum_{\nu}
\qquad
a_3 \cuum_{\nu} = \lambda_3 \cuum_{\nu}
\qquad
\pp \cuum_{\nu} = \nu \cuum_{\nu}.
\end{eqnarray}
From eq.\ \rf{jz2} it follows that if we choose  
\begin{eqnarray}\label{jz3}
\lambda_1 = -\frac{\mu}{2g\sqrt{\GG}}
\qquad
\lambda_3 = \frac{1}{6 g\sqrt{\GG}}
\qquad
\nu = \frac{1}{\sqrt{\GG}}
\end{eqnarray}
and make the identification 
\beq\label{jz4}
a_n \to V(\lam_1,\lam_3) \, a_n V^{-1}(\lam_1,\lam_3) = 
a_n - \lambda_1 \delta_{n,1} - \lambda_3 \delta_{n,3}
:= \sqrt{\GG}\, \phi_n
\eeq
\beq\label{jz4a}
a_n^\dagger \to V(\lam_1,\lam_3) \, a_n^\dg V^{-1}(\lam_1,\lam_3)
:= \frac{1}{\sqrt{\GG}}\, \phi_n^\dagger
\eeq
then eqs. \rf{js1} and \rf{jy2} become consistent with \rf{jz6}. 
We can finally write 
\begin{eqnarray}\label{jz5}
g \sqrt{\GG}\, { \hH}_{\rm W}\Big|_{p=1/\sqrt{G}}
\ &=& \ 
{\hH}
\,-\,
  \frac{1}{\GG} \bigg(
  \frac{\mu^2}{4 g}
+ \frac{1}{4 g} \phi_4^\dagger
- \frac{\mu}{2 g} \phi_2^\dagger
+ \phi_1^\dagger
\!\;\bigg).
\end{eqnarray}
valid on the subspace of $\cH$ where the eigenvalue of $p$ is 
$1/\sqrt{G}$. This is our basic relation. By acting with the vertex operator 
$V(\lam_1,\lam_3)$ defined in \rf{jz1} 
on the absolute vacuum $|0\ra$ we create 
a condensation of $\phi_1^\dg$, $\phi^\dg_3$ and $q$ modes. This condensate
defines the coupling constants of a GCDT string field theory, 
but if our starting point is  $\hH_W$ the corresponding $GCDT$ 
vacuum $\cuum_{\nu}$ is unstable, as is clear from \rf{jz5}.

It is the condensation of $\phi_3^\dg$ which creates a non-zero $\lambda_3$ 
and it is this non-zero $\lam_3$ which results in  the appearance of the term 
$- \sum_{l=1}^\infty \phi_{l+1}^\dagger l \phi_l$. Such a term is necessary 
if we want to have the possibility of an expanding universe. In the 
physical vacuum $\cuum$ the universe can thus both expand and 
contract and the parameter $T$ multiplying the Hamiltonian can 
then be interpreted as the time-evolution parameter of the universe.   
One can say that time $T$ refers to a vacuum $\cuum_{\nu}$ and only allows 
for an interpretation as the cosmological time of a  spacetime 
after $\cuum_{\nu}$ is introduced.

\section{Discussion}\label{discuss}

We have attempted to create a model of the universe where
there is an ``absolute'' vacuum $|0\ra$ and a ``pre-geometry'' 
Hamiltonian $\hH_W$. We were inspired by non-critical 
string field theory to choose the simplest possible 
non-trivial $\hH_W$, related to the $W^{(3)}$ algebra. 
The corresponding partition function \rf{jy4} 
can most likely be related to a tau-functions of a KP hierarchy 
(details are being worked out), but as mentioned the system does not offer an 
obvious interpretation as a dynamical system for spacetime. 
However, acting with a vertex operator on the absolute vacuum
brings us to a coherent state \rf{jz1a}, $\cuum_{\nu}$,
which has non-zero overlap to the absolute vacuum. We denote $\cuum_{\nu}$
a ``physical'' vacuum because the corresponding action \rf{jz4}
on creation and annihilation operators, which amounts to a simple shift
of expectation values of the operators in $\hH_W$, leads to  an 
interpretation  of $\hH_W$ as a Hamiltonian which creates, annihilates
and changes space, thus creating a dynamical spacetime, relative
to this physical vacuum. At the same time the simple shifts of 
expectation values define the coupling constants of the string
field Hamiltonian which governs the evolution. Clearly this process
has some resemblance to standard spontaneous symmetry breaking 
where  the vacuum expectation values of a field might define the 
values of some of the coupling constants of the theory. At the same
time this ``symmetry breaking'' becomes the source of a ``Big Bang'',
the creation of a universe from nothing since 
$\hH_W|_{p=1/\sqrt{G}}$ contains the creation operators 
which will act non-trivially on $\cuum_{\nu}$. 
Once the choice of $\cuum_{\nu}$ is 
made $T$ can be viewed as a cosmological time and space can 
next be created due to the instability of $\cuum_{\nu}$ with respect to $\hH_W$.
The origins of space and time are thus different in our model, time being
a ``precursor'' for space, a point
also emphasized in \cite{cs} although from a different perspective.
Many universes can be created and they can join and split as a function
of $T$ and we can explicitly calculate such amplitudes \cite{GSFT}. 
Let $\cH(\lam_1,\lam_3,\nu)$ be the Fock space spanned by states
obtained by acting  repeatedly with the $\phi_n^\dg$ operators on
$\cuum_{\nu}$. In the larger Hilbert space $\cH$ of 
$\hH_W$ we have that
\beq\label{ju1}
\cH(\lam'_1,\lam'_3,\nu')\; \bot\; \cH(\lam_1,\lam_3,\nu)~~~{\rm for}~~~
\nu'\neq \nu.
\eeq
since the operator $p$ is Hermitian. However, 
all Hilbert spaces with the same value of $\nu$ but 
different values of $\lam_1$ and $\lam_3$ are identical since 
the overlaps between different coherent states created by acting with 
$V(\lam_1,\lam_3)$ for different values of $\lam_1$ and $\lam_3$ are 
non-zero. Thus universes with different coupling constants
can in principle interact if we can provide a suitable interaction
term and this interaction could  change the 
values of the coupling constants of the universes. One could 
call such a scenario a ``fourth quantization'' since our 
string field theory is already a ``third quantization'' as mentioned above.
One could imagine to use such change in 
coupling constants to explain aspects of inflation, provided
suitable higer-dimensional models can be consistently formulated
\cite{toappear}.

This brings us to a missing ingredient in our construction,
namely a mechanism for choosing a specific physical vacuum $\cuum_{\nu}$.
Being minimalistic one could say that the probability $P(\lam_1,\lam_2)$
of being in a universe corresponding to a given choice of cosmological 
constant and a given choice of coupling constant $g$ would be 
given related to the overlap between $|0\ra$ and $\cuum_{\nu}$, i.e.
\beq\label{jf1}
P(\lam_1,\lam_3) \propto \Big| \la 0 \cuum_{\nu}\Big|^2 \propto 
\e^{-\lam_1^2-\lam_3^2}.
\eeq
where the relation between coupling constants and $\lam_1$ and 
$\lam_3$ is given by eqs.\ \rf{jz3},
but it would be desirable to have a dynamical mechanism for selecting
$\cuum_{\nu}$. Also, a statement like \rf{jf1} does not make 
much sense if one allows interactions between universes with 
different coupling constants.

It would be interesting to generalize the model to include 
matter, in particular in such a way that the choice of physical vacuum
$\cuum$ would not only be a choice of the coupling 
constants related to geometry
but also a choice of matter content. Understanding the mechanism for 
the choice of such $\cuum$ would be exciting. Equally exciting is the 
possibility to extend the considerations here to genuine four-dimensional 
models. All this indeed seems possible \cite{toappear}.    

\vspace{1cm}

\noindent {\bf Acknowledgments.} 

JA and YW acknowledge support from the ERC-Advance grant 291092,
``Exploring the Quantum Universe'' (EQU). 
JA  were supported in part by Perimeter Institute of Theoretical Physics.
Research at Perimeter Institute is supported by the Government of Canada
through Industry Canada and by the Province of Ontario through the 
Ministry of Economic Development \& Innovation.


\end{document}